\documentclass{iaus}
\usepackage{graphicx}

\newcommand{\tef}{T_\mathrm{eff}}
\newcommand{\logg}{\log g}

\title[Is the Sun typical?] 
{The Sun. A typical star in the solar neighborhood?}

\author[Jorge Mel\'endez]   
{Jorge Mel\'endez$^1$
}

\affiliation{$^1$Departamento de Astronomia do IAG/USP, Universidade de S\~ao Paulo, Rua do Mat\~ao 1226, Cidade Universit\'aria, 
05508-900 S\~ao Paulo, SP, Brazil. \\ email: {\tt jorge@astro.iag.usp.br} \\[\affilskip]
}

\pubyear{2013}
\volume{298}  
\pagerange{1--6}
\jname{Setting the scene for Gaia and LAMOST - the current and next generations of surveys and models}
\editors{A.C. Editor, B.D. Editor \& C.E. Editor, eds.}
\begin{document}

\maketitle

\begin{abstract}
The Sun is used as the fundamental standard in chemical abundance studies, thus it is important to know whether 
the solar abundance pattern is representative of the solar neighborhood. Albeit at low precision (0.05 - 0.10 dex) 
the Sun seems to be a typical solar-metallicity disk star, at high precision (0.01 dex) its abundance pattern seems 
abnormal when compared to solar twins. The Sun shows a deficiency of refractory elements that could be due to 
the formation of terrestrial planets. The formation of giant planets may also introduce a signature in 
the chemical composition of stars. We discuss both planet signatures and also the enhancement of neutron-capture elements 
in the solar twin 18 Sco.

\end{abstract}

\firstsection 
\section{Introduction}

In previous studies the question of whether the Sun has a normal composition has been discussed in detail 
(Gustafsson 1998, 2008; Gustafsson et al. 2010; Allende Prieto 2008, 2010). It has been recognized that earlier 
studies obtained discrepant results regarding possible chemical abundance anomalies in the Sun, probably due 
to relatively large (0.05 - 0.10 dex) abundance uncertainties. An illustrative example is the conflicting 
results found for the [O/Fe] ratio in solar-metallicity thin disk dwarfs. In the seminal study by Edvardsson et al. (1993), 
the [O/Fe] ratio around [Fe/H] = 0 in disk stars was found to be somewhat subsolar, meaning that the Sun could be 
either somewhat oxygen-rich or somewhat iron-poor, relative to disk stars. Different results were obtained by 
Bensby et al. (2004) and Ram\'{i}rez et al. (2007), who found that solar-metallicity disk stars have [O/Fe] $\sim$ 0 
and [O/Fe] $\sim$ +0.1 dex, respectively. These three different studies show that depending on how the analysis is made, 
worrisome differences of up to 0.15-0.20 dex could be found in the [O/Fe] ratios of disk stars. 
Even the consistent analysis of three different oxygen 
abundance indicators by Ecuvillon et al. (2006), resulted in discrepant oxygen abundances 
(solar, sub-solar and super-solar) regarding the [O/Fe] ratios of disk stars. 

A serious issue when comparing the Sun to other solar-type stars is the effective temperature scale, as 
different photometric $\tef$ scales (e.g., Alonso et al. 1996; Ram\'{i}rez \& Mel\'endez 2005; 
Casagrande et al. 2010; Boyajian et al. 2013) may have different zero-points. On this regard, it is important 
to check the zero-points of a given temperature scale using solar colors as those derived recently by 
Mel\'endez et al. (2010), Ram\'{i}rez et al. (2012) and Casagrande et al. (2012), using a large sample of solar twins 
and solar analogs. The same concern applies to photometric metallicities, i.e., their zero-point must be tested 
before they are used to compare the Sun to other stars (e.g., \'Arnad\'ottir et al. 2010; Mel\'endez et al. 2010; Datson et al. 2012).

Another problem that may arise when comparing the Sun to other stars is the adopted line list and atomic data, 
and how that data is used to obtain spectroscopic stellar parameters ($\tef, \logg, v_t$). Sometimes  
gf-values are ``calibrated'' using very high resolution solar atlases, and then applied to stars 
that have been observed using a spectrograph with a much lower resolution, where the line profiles are not as well 
defined as in the solar atlas. Even worse would be to use solar gf-values determined by a different group 
using a different set of model atmospheres. The use of absolute (laboratory or theoretical) gf-values is also 
problematic, as their large uncertainties could imply relatively large errors (low precision) in the obtained stellar parameters. 
Also, although naively we would expect that absolute gf-vaues would result in accurate spectroscopic
stellar parameters, this is not necessarily the case due to problems in model atmospheres (Asplund 2005) 
and the simplifications made for the treatment of line formation (LTE vs. NLTE; e.g., Lind, Bergemann \& Asplund 2012). 
Also, precise absolute gf-values exist only for a minority of the lines, meaning that to obtain spectroscopic equilibrium
using FeI and FeII lines we may have to use some imprecise absolute gf-values. The same applies to the
determination of chemical abundances for elements other than iron. 
It is also dangerous to use different sets of lines for the Sun and for the stars, because due to the 
large errors in the absolute gf-values, a different set of lines could lead to a different solution, 
resulting in inaccurate and imprecise stellar parameters and stellar abundances. 

Blends are also a serious issue, as the dependence of a given line to both stellar parameters and 
abundance of the main component could be affected by the minor component of the blend. This problem 
could be minimized by selecting clean lines or lines with weak blends, or also by deblending or 
spectrum synthesis. Another issue is both continuum determination during the data reduction and 
continuum placement in the measurements. The Sun and the sample stars must have a similar determination 
of the continuum. The way the measurements are done is also very important.  It is increasingly common 
in the literature to use automatic tools such as ARES (Sousa et al. 2007) to measure the equivalent widths, 
yet the measurements done for a given sample star may not be identical to the way the measurement 
is done for the Sun. Even solar twins that may look almost identical to the Sun can show small variations 
in their spectra relative to the solar spectrum, due to somewhat different stellar parameters (Teff, log g, vt), 
different rotation velocities, different chemical compositions and different levels of telluric contamination. 
For a given line, ARES does a continuum normalization in a small spectral window, and due to the small 
variations mentioned above, the continuum fit done by ARES may not be necessarily the same for the stars 
and the Sun. Also, ARES may use a different number of components (lines) to model a given spectral window, 
and that number may vary from star to star. Although the automatic measurements are good enough for a 
typical abundance analysis, they may not be adequate to distinguish tiny variations (0.01 dex) between 
the chemical compositions of stars and the Sun.

In summary, systematic differences between the stars and the Sun could arise due to the 
(i) analysis techniques (equivalent widths vs. spectrum synthesis),  
(ii) stellar parameters, 
(iii) adopted grid of model atmospheres, 
(iv) treatment of line formation (LTE vs. NLTE), 
(v) adopted gf-values, 
(vi) adopted line lists, 
(vii) spectral resolution, 
(viii) signal-to-noise ratio, 
(ix) problems with the spectrograph, 
(x) adopted solar spectrum (sky, Moon, moons of other planets, asteroids, solar atlas), 
(xi) data reduction, 
(xii) determination of the continuum, 
(xiii) blends, 
(xiv) equivalent width measurements, and
(xv) adopted solar abundances (e.g., Asplund et al. 2009), when solar abundances are not determined in the same analysis. 

Many of the above problems could be eliminated or largely minimized by planning the observations 
of the stars and the Sun using the same spectrograph and the same setup. Even better would be 
to observe the Sun in the same observing run as the stars, to guarantee that there were 
no significant changes in the instrument configuration. The solar spectrum could be obtained 
using asteroids, which are almost point sources, so that the data reduction is performed 
in the same way as for the stars. Continuum normalization should be performed homogeneously 
in the stars and the Sun. Having obtained the stellar and solar spectra in the same way, 
we can now perform a strictly line-by-line differential analysis, so that we make sure that 
every line in the spectrum is measured in the same way in the stars and the Sun, using the 
same continuum regions and using exactly the same part of the line profile. Then, the abundances 
can be determined line-by-line, so that uncertainties in the gf-values will cancel out. 
It would be even better to perform this kind of analysis using solar twins, stars with spectra 
very similar to the Sun (e.g., Mel\'endez et al. 2006, 2009; Ram\'{i}rez et al. 2009;
Takeda \& Tajitsu 2009; Datson et al. 2012).
In this way, many of the modeling uncertainties mentioned above are minimized, 
resulting in extremely precise abundances at the 0.01-0.02 dex level, provided very high resolution
(R $\geq$ 60000) very high S/N ($\geq$200) spectra are employed. This is the approach 
first adopted by Mel\'endez et al. (2009) and Ram\'{i}rez et al. (2009), that led to the discovery 
of abundances anomalies in the chemical composition of the Sun. The Sun seems chemically anomalous 
when compared to nearby solar twins in the thin disk, perhaps due to the formation of terrestrial 
planets in the Sun. The abundances anomalies that may be caused by planet formation are discussed 
in more detail below.

\section{The stellar chemical abundance - planet connection}

The processes of star and planet formation are far from fully understood, but we know that they 
occur nearly at the same time. Thus, it is likely that the formation of both terrestrial 
and gas giant planets has imprinted signatures in the chemical composition of their host stars. 
During the final phases of gas accretion by the star, when the proto-planetary nebula cools, 
dust grain condensates and later coagulates to form planetesimals and finally terrestrial planets 
in the inner region, while in the outer region rocky-ice embryos will accrete hydrogen and helium 
to form the gas giant planets (\cite[Morbidelli et al. 2012]{mor12}), although giant planets could also form by gravitational 
disk instability. The chemical fingerprints left by these processes are key to constrain different 
models of planet formation. The classical example of the star-planet connection is the well-known 
planet-metallicity correlation in solar type stars, indicating that a higher abundance of metals increases 
the probability of forming giant planets (\cite{gon97} 1997). This signature has been confirmed 
by further works (e.g. \cite{fv05} 2005; \cite{us07} 2007; \cite{ghe10} 2010) 
and may favor the theory of core-accretion for giant planet formation.

Although the correlation between giant planet frequency and overall stellar metallicity has been important 
in the context of planet formation, it is probably only the tip of the iceberg regarding the connection 
between stars and planets. In the subsequent decade after the seminal work by \cite{gon97} (1997), there have 
been many attempts to explore other relations between planets and the abundance of specific chemical 
elements, but no clear result was obtained (\cite{us07} 2007), probably due to the relatively large uncertainties 
(about 0.05 dex) in the determination of chemical abundances (\cite{asp05} 2005), or in same 
cases due to a bias in the comparison between stars with and without planets, 
as shown for example for lithium, for which a fare comparison found no difference 
in the Li abundances of the two samples (\cite{bau10} 2010). 

As mentioned above, errors as low as 0.01 dex could be obtained in a 
strictly differential analysis of a sample of stars that are very similar 
to each other (due to the cancellation of systematic errors), opening thus new windows 
on the planet-star connection. Below we will discuss recent works 
on high precision chemical abundances that are giving further clues on the formation of rocky 
(\cite{mel09} 2009) and giant (\cite{ram11} 2011) planets.

\begin{figure}[b]
\begin{center}
 \includegraphics[width=5.4in]{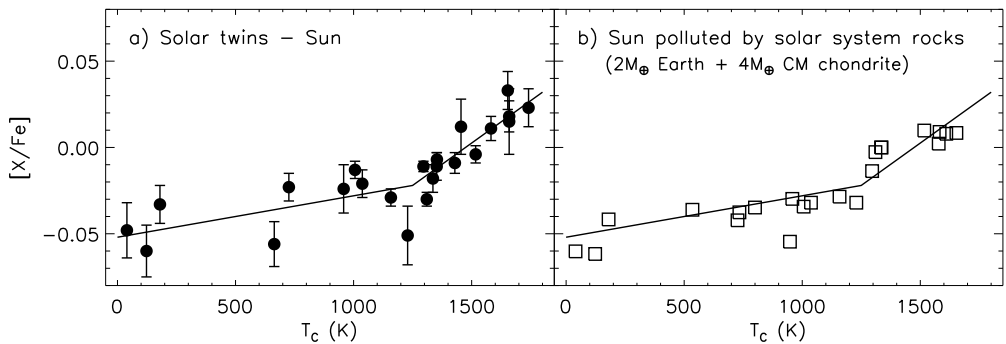} 
 \caption{\small {\bf a)} Average elemental abundance ratios (relative to Fe) as a function 
of condensation temperature for 11 solar twin stars from \cite{mel09} (2009). This clear correlation 
between [X/Fe] and $T_\mathrm{C}$ was not detected before due to the large errors in standard 
chemical composition analysis ($\sim\pm0.05$\,dex). In \cite{mel09} (2009), 
the [X/Fe] values have errors of about 0.01\,dex. The solid line is a linear fit to the data, 
but broken at $T_\mathrm{C}=1250$\,K. {\bf b)} Variation of the solar photospheric abundances, 
relative to Fe, if the Sun's present-day convective envelope were polluted by 2 Earth masses 
of Earth-like material and 4 Earth masses of CM chondrite rocks. Note the excellent agreement 
with the linear fit to the solar twin chemical abundance data (the solid line in this panel is the same as in panel a).
Adapted from \cite{mel12} (2012).}
   \label{fig_sun}
\end{center}
\end{figure}

\section{Signatures of terrestrial planet formation}

A new era on precise chemical abundance determinations started with the analysis of 
11 solar twins by \cite{mel09} (2009). As these stars have both stellar parameters 
and spectra nearly indistinguishable from the Sun, many systematic errors that plague 
stellar abundance analyses are canceled in a strictly line-by-line differential analysis of the 
solar twins relative to the Sun, being possible to achieve an unprecedented precision 
of 0.01 dex (\cite{mel09} 2009; \cite{mel12} 2012), about a factor 
of 5 smaller than standard abundance analyses. This dramatic improvement is critical 
to explore the small effects that planet formation may imprint on the chemical composition of stars. 

\cite{mel09} (2009) found that the Sun is peculiar 
when compared to a sample of 11 solar twins (Fig. 1a). The abundance of refractory elements 
(those that condensed at high temperatures in the inner solar nebula) is systematically 
smaller in the Sun relative to the solar twins. The amount of material missing is compatible 
with the amount needed to form the terrestrial planets (\cite{mel09} 2009; \cite{gus10} 2010; \cite{cha10} 2010), 
meaning that the deficiency of refractory elements in the solar photosphere would disappear 
if the rocky material formed in the solar system were diluted into the present-day solar 
convective envelope (\cite{cha10} 2010; \cite{mel12} 2012), as shown in Fig. 1b. This opens 
the exciting possibility of discovering stars that could potentially host terrestrial planets 
based on a careful chemical abundance analysis, but certainly further work is needed to consider 
this as a firm signature of rocky planets.

If the above signature is confirmed, then stars with chemical composition similar to 
solar could potentially host terrestrial planets. So far the stars that have their chemical composition 
most similar to the Sun are the solar twins HIP 56948 (\cite{mel07} 2007; \cite{mel12} 2012) and 
HIP 102152 (Monroe et al. 2013). Interestingly, 
the precise radial velocity observations obtained for HIP 56948 at the McDonald and Keck 
observatories, and for HIP 102152 at La Silla, show no signs of giant planets within and in 
the habitable zone, making these stars excellent candidates for hosting habitable rocky planets.

\begin{figure}[b]
\center{\includegraphics[width=2.1in]{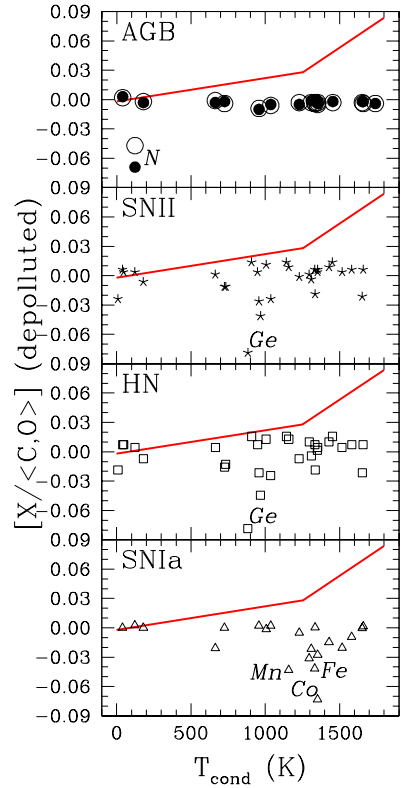}}
\caption{
Abundance ratios obtained after de-polluting the solar nebula from
contamination by an AGB star (circles), 
SNII (stars), hypernova (squares) and SNIa (triangles).
In the top panel the effect of adopting different solar abundances 
(open circles: \cite{and89} 1989; filled circles: \cite{asp09} 2009) is shown.
The solid line represents the mean abundance pattern
of 11 solar twins relative to the Sun (\cite{mel09} 2009). None of the
pollution scenarios can explain the trend with condensation temperature.
The chemical elements that change the most are labeled. Figure taken from \cite{mel12} (2012).
}
\label{pollution}
\end{figure}

\section{Is there any other explanation for the Sun's abundance anomalies?}

As discussed extensively in Mel\'endez et al. (2009, 2012), the abundances anomalies 
cannot be explained by other causes such as contamination from AGB stars, SNIa, SNII or HN (Fig. 2), 
or by Galactic chemical evolution processes or age effects. 
\cite{kis11} (2011) have shown that the peculiar abundance pattern 
cannot be attributed to line-of-sight inclination effects. Also, the abundance trend 
does not arise due to the particular reflection properties of asteroids. 
Although the abundance peculiarities may indicate that the Sun was born in a massive 
open cluster like M67 (\cite{one11} 2011), this explanation is based on the analysis of only 
one solar twin. 

So far the best explanation for the abundance trend seems to be the formation of 
terrestrial planets. The Kepler mission should detect the first Earth-sized
planets in the habitable zones of solar type stars. We look forward
to using 8-10 m telescopes to perform careful differential
abundance analyses of those stars, in order to verify if our chemical
signatures indeed imply rocky planets.

\begin{figure}[b]
\begin{center}
 \includegraphics[width=3.5in]{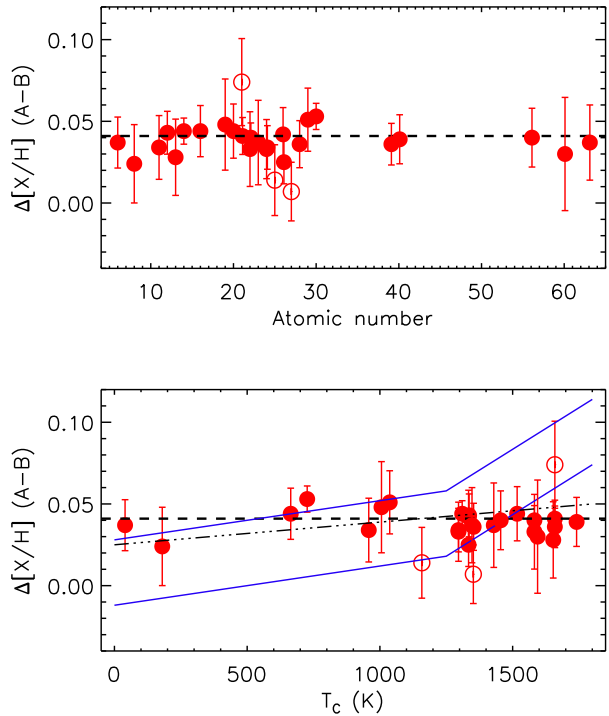} 
 \caption{Top panel: elemental abundance difference between 16 Cyg A and B as a function of atomic number. 
Open symbols show the three species more discrepant from the mean: Sc\,\textsc{i} (21), Mn (25), and Co (27). 
Bottom panel: as in the top panel for the abundance differences versus dust condensation temperature. 
The dashed line is at +0.041 while the solid lines represent the mean trend of solar twins 
by \cite{mel09} (2009) with two arbitrary offsets. The dot-dashed line corresponds to 
a slope of $1.4\times10^{-5}$\,dex\,K$^{-1}$, as derived by \cite{law01}.}
   \label{fig3}
\end{center}
\end{figure}

\section{New signatures of giant planet formation}

Regarding the formation of giant planets, we have recently studied the pair 
of solar analogs 16 Cyg A and B. This pair is very important to understand 
more about giant planet formation, because the star 16 Cyg B hosts a gas 
giant planet at 1.7 A.U. (\cite{coc97} 1997), while no planets have been 
detected yet around 16 Cyg A. The binary pair is supposed to have the same 
chemical composition, as it was formed from the same natal cloud, unless the 
formation of the giant planet around 16 Cyg B altered its chemical abundances. 

Our careful differential abundance analysis between 16 Cyg A and B showed that 
the component B is systematically more metal-poor, by about 10\% ($\sim$0.04 dex), 
in the two dozen chemical elements analyzed (\cite{ram11} 2011), 
as shown in Fig. 3. Thus, it seems that the formation of the gas giant around 16 Cyg B 
robbed a fraction of the metals present in its parent nebula (\cite{ram11} 2011). 
Nevertheless, another abundance analysis (at somewhat lower precision) published 
nearly at the same date, showed no difference between the abundance pattern of 
components A and B (\cite{sch11} 2011). Although the spectra used 
by \cite{ram11} (2011) have a higher resolving power and a higher precision was 
achieved, it would be important to perform a new abundance analysis of the pair 
to confirm the signature of giant planet formation.

\begin{figure}[b]
\begin{center}
 \includegraphics[width=3in]{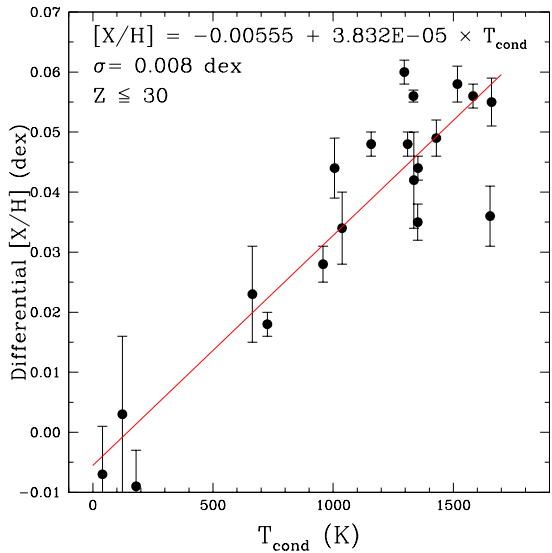} 
 \caption{Trend with dust condensation temperature for the solar twin 18 Sco (Mel\'endez et al., in preparation).
The element-to-element scatter is only 0.008 dex.}
   \label{fig4}
\end{center}
\end{figure}

\section{Planet search around solar twins}

The above signatures of terrestrial and giant planets are telling us
that there could be a close connection between stellar chemical composition and
planet architecture. Solar twins are ideal to obtain very precise stellar
abundances but unfortunately planet information is lacking for most of
them, so currently we cannot study in detail the relation between
chemical abundance anomalies and different type of planets. 

The synergy between the high precision (0.01 dex) chemical abundances 
obtained in solar twins (Fig. 4) and high precision (1m s$^{-1}$) radial velocities
that can be obtained with HARPS, 
can give us new insights into the planet-stellar connection. In order to study 
with unprecedented detail the connection between chemical abundances and planet 
architecture, we have been granted 88 nights with HARPS for a Large ESO Programme 
that started in October 2011 and should continue until 2015. This in an international 
collaboration involving astronomers from Brazil, Germany, USA and Australia.

Around 70 solar twins are being observed at the ESO La Silla observatory
and already some of them are showing radial velocity variations compatible 
with planets. For all the sample stars we have acquired 
high resolution high S/N spectra using the MIKE spectrograph at the 6.5 m Magellan 
telescope in Las Campanas, in order to obtain a homogeneous set of high precision 
(0.01 dex) chemical abundances. Also, some stars have spectra of even
better quality acquired with UVES at the VLT. 

Our initial chemical abundance analyses show that on a star-by-star basis we can obtain 
chemical abundances with a precision at the 0.004 - 0.008 dex level, i.e., even better 
than initially anticipated (0.01 dex). For the solar twin HIP 56948, 
the element-to-element scatter is only 0.004 dex for the volatile elements and 0.008 dex 
for the refractories (\cite{mel12} 2012), and our on-going work on the solar twin 18 Sco 
shows also an element-to-element scatter of only 0.008 dex (Fig 4). Similar results are
being obtained for other solar twins (Monroe et al. 2013).

\begin{figure}[t]
\begin{center}
 \includegraphics[width=3in]{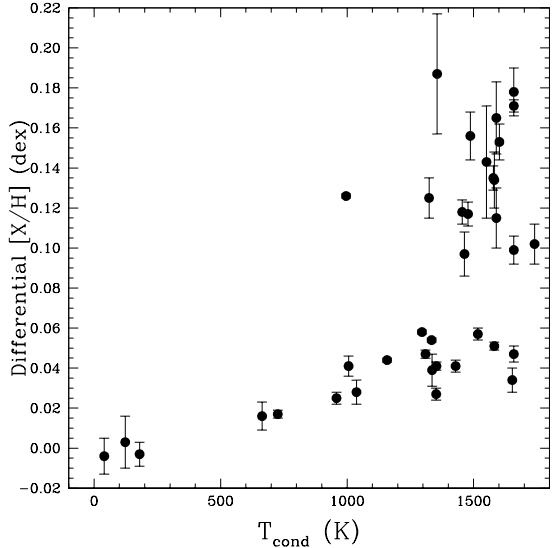} 
 \caption{Complex abundance pattern of light and heavy elements in the solar twin 18 Sco (Mel\'endez et al., in preparation).
Besides the trend with condensation temperature shown in Fig. 4, all n-capture elements show a further enhancement 
([X/H] $>$ 0.09 dex)}
   \label{fig5}
\end{center}
\end{figure}

\section{Signatures of neutron-capture elements in the solar twin 18 Sco}

The solar twin 18 Sco presents a complex abundance pattern (see Fig. 5). On top of the typical 
trend with condensation temperature seen in other solar twins (Mel\'endez et al. 2013; Monroe et al. 2013),
this star also shows a large enhancement of the neutron-capture elements. 
Using yields of AGB stars computed by Amanda Karakas, we can verify if the observed
enhancement is due to AGB pollution. After subtracting both the trend with condensation
temperature and the s-process contribution from AGB stars, we are left with an abundance
signature identical to the r-process in the solar system (Figure 6). This specific abundance pattern
can only be obtained thanks to the high precision (0.01-0.02 dex) we achieved for these heavy elements.

\begin{figure}[t]
\begin{center}
 \includegraphics[width=3in]{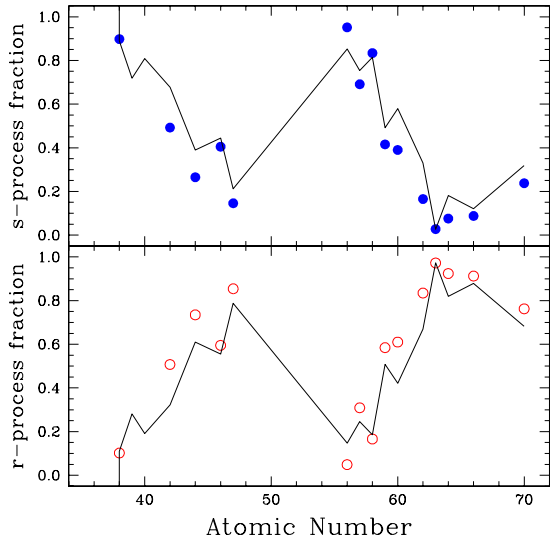} 
 \caption{Abundance pattern of the n-capture elements in 18 Sco. 
In the upper panel the abundance pattern due to the AGB contribution (filled circles)
is compared to the s-process in the solar system (line).
The residual abundances in 18 Sco after subtracting the trend with condensation 
temperature and the s-process from AGB stars, are shown by open circles in the lower panel. This
abundance pattern is in excellent agreement with the r-process in the solar system (line).}
   \label{fig6}
\end{center}
\end{figure}

\section{Conclusions}
Our work has shown that it is possible to obtain chemical abundances with a precision of 
about 0.01 dex or even better ($\sim$0.005 dex) in some cases (\cite{mel12} 2012). 
This abundance precision can be achieved in a strictly differential line-by-line analysis.
Our approach minimizes systematic errors due to deficiencies in the adopted model atmospheres.
This is shown in Fig. 7, where the difference in differential abundances obtained using
two dissimilar grids of model atmospheres (MAFAGS models vs. Kurucz overshooting models;
see Mel\'endez et al. 2012) is shown. 
The average difference is only 0.001 dex, with an element-to-element scatter of only 0.00075 dex.
The unprecedented precision that we achieve ($\sim$0.01 dex) was key to discover that the Sun is not a typical star.
The abundance anomalies in the Sun may be connected to the formation of terrestrial planets 
(\cite{mel09} 2009; \cite{ram09} 2009, 2010). Also, giant planet formation may imprint a
signature in the chemical composition of stars (\cite{ram11} 2011). Our on-going planet search around 
solar twins will allow us to study at unparalleled precision any connection that 
may exist between planet architecture and chemical abundance peculiarities.
Finally, our high precision abundances can give us better insights on the neutron-capture elements,
which could be important for chemical tagging in our Galaxy.

\begin{figure}[t]
\begin{center}
 \includegraphics[width=3in]{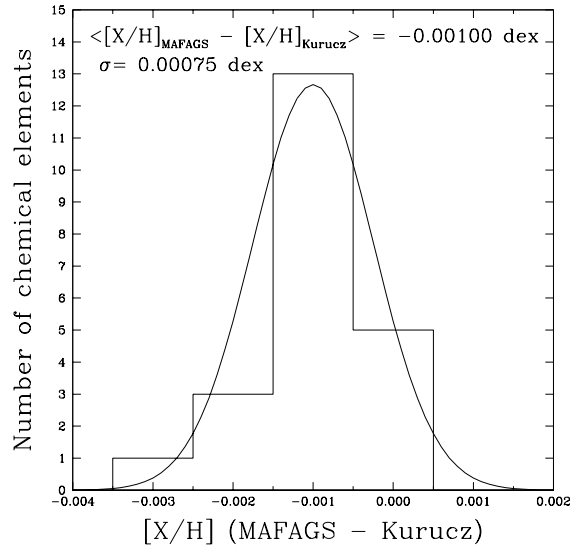} 
\caption{
Histogram of the differences between differential chemical abundances 
obtained with MAFAGS models and Kurucz models. The element-to-element
scatter is only $\sigma$ = 0.00075 dex, i.e., 0.17\%.
}
   \label{fig7}
\end{center}
\end{figure}

\section{Acknowledgments}
J.M. thanks support from FAPESP (2012/24392-2). This work is based on observations
obtained at Keck, Magellan/Las Campanas, McDonald, VLT/Paranal (083.D-0871)
and La Silla (188.C-0265).

\begin{discussion}

\discuss{Xiaowei Liu}{(1) With the current radial velocity measurement accuracy, 
I assume it would be extremely difficult to rule out the presence of rocky planets in stars of 
your solar twin sample. (2) Assuming 50\% of solar-type stars harbor rocky planets, 
do you see some of your solar twin stars showing exactly the same abundance pattern? 
(3) Would the abundance pattern amongst solar twins actually tell us more about planetary systems in those stars?}

\discuss{Jorge Mel\'endez}{(1) Yes, rocky planets similar to the Earth cannot be detected with current instruments. 
However, we can use planets detected using the transit method by the Kepler mission. 
There is a large number of Kepler planet candidates with a radius as small as the Earth, 
and presumably a good fraction of them may be rocky planets. We plan to obtain chemical abundances 
of the host stars of those potential rocky planets, to verify our terrestrial planet hypothesis. 
(2) Yes, a fraction of the solar twins shows an abundance pattern similar to the Sun. 
We have found that about 15-20\% of solar twins have the same chemistry as the Sun within 
the abundance errors. (3) This is exactly the goal of our 4-year (2011-2015) planet survey 
around solar twins with HARPS, to see if different planetary architectures could be associated 
to different abundance patterns. The analysis of the first star in our HARPS sample 
(Monroe, Mel\'endez, Ram\'{i}rez et al. 2013), HIP 102152, shows that this star has 
a solar abundance pattern. Interestingly, the HARPS observations performed so far shows 
that this star does not have giant planets in its inner region. Therefore, so far are similarities 
to the Sun both in its abundance pattern and in the lack of inner giant planets.}

\discuss{Livia Origlia}{In order to get such level of accuracy in the abundance estimates and 
high S/N you also used very precise flat-fielding of the observed spectra. Can you please comment on that?}

\discuss{Jorge Mel\'endez}{The main problem related to flat-fielding was the fringing seen in the UVES 
spectra that were reduced by its pipeline. So, to improve the results the data was fully reduced by hand. 
Notice that the spectra falls on different pixels, so we do not need to achieve very high S/N in a given 
pixel of the combined flat field. Also, in some cases multiple exposures were combined to achieve a higher S/N. 
This is the case of the Li region, for which we combined a standard with a non-standard setup, 
so that we could have some overlap between both setups around the LiI 670.8nm line.} 

\discuss{Xiaoting Fu}{You mentioned that the Sun abundance is peculiar because it uses the 
material to form the dust and planets. Have you checked the abundance of the planet-host stars in Kepler.}

\discuss{Jorge Mel\'endez}{We intend to verify the abundance pattern of the host stars of Kepler 
planet candidates with radius similar to Earth's, i.e., potential rocky planets. 
We have an observing run with Keck/HIRES in August 2013 to perform this work.}

\discuss{Hans-Walter Rix}{You called the Sun ``peculiar'' in the comparison to the 11 solar twins. 
Is any star's abundance pattern ``normal'' when looked at with this precision? 
I.e., does the Sun's abundance pattern really deviate much more from the other stars 
than any of the other 11 stars deviate from the average?}

\discuss{Jorge Mel\'endez}{It is hard to define a ``normal'' abundance pattern. 
It could be defined either by the mean (in Mel\'endez et al. 2009 the trimean was adopted) 
chemical composition of the solar twins, or by the most refractory-rich stars (i.e. those stars 
who have the least depletion of refractory elements, as defined recently by Ram\'{i}rez et al. 2013). 
In both definitions the Sun seems abnormal at a significant level. In Mel\'endez et al. (2013) 
we found two solar twins with large deviations from the mean abundance pattern of the twins, i.e., 
with chemical compositions similar to the Sun. We are analyzing larger samples of solar twins to 
better quantify the fraction of stars with chemical compositions similar to the Sun.} 

\discuss{Chiaki Kobayashi}{For 18 Sco, what a beautiful agreement with the solar s- and r-process! 
How did you subtract the AGB contribution.}

\discuss{Jorge Mel\'endez}{We used AGB yields computed by Amanda Karakas and diluted 
those yields into a one-solar-mass nebula, then the results were subtracted from the 
observed pattern of the neutron-capture elements in 18 Sco. Amanda please could you give us more details about it?}

\discuss{Amanda Karakas}{Comment: I took a model of a 3 M$_\odot$ AGB stars of Z = 0.01 and diluted 
the yields (1\% of mass of material injected) into a 1 M$_\odot$ proto-solar cloud of solar composition. 
The nucleosynthesis model includes the s-process elements.}

\end{discussion}


\begin{thebibliography}{}
\bibitem[Allende Prieto(2008)]{2008mru..conf...30A} Allende Prieto, C.\ 
2008, \textit{The Metal-Rich Universe}, 30 

\bibitem[Allende Prieto(2010)]{2010IAUS..265..304A} Allende Prieto, C.\ 
2010, \textit{IAU Symp.}, 265, 304

\bibitem[Alonso et 
al.(1996)]{1996A&A...313..873A} Alonso, A., Arribas, S., \& Martinez-Roger, C.\ 1996, \textit{A\&A}, 313, 873 

\bibitem[Anders \& Grevesse]{and89} Anders, E., \& Grevesse, N.\ 1989, \textit{Geochim. Cosmochim. Acta}, 53, 197 

\bibitem[{\'A}rnad{\'o}ttir et 
al.(2010)]{2010A&A...521A..40A} {\'A}rnad{\'o}ttir, A.~S., Feltzing, S., \& Lundstr{\"o}m, I.\ 2010, \textit{A\&A}, 521, A40 
\bibitem[Asplund]{asp05} Asplund, M.\ 2005, \textit{ARAA}, 43, 481 

\bibitem[Asplund et al.]{asp09} Asplund, M., Grevesse, N., Sauval, A.~J., \& Scott, P.\ 2009, \textit{ARAA}, 47, 481 

\bibitem[Baumann et al.]{bau10} Baumann, P., Ram{\'{\i}}rez, I., Mel{\'e}ndez, J., Asplund, M., \& Lind, K.\ 2010, \textit{A\&A}, 519, A87 

\bibitem[Bensby et 
al.(2004)]{2004A&A...415..155B} Bensby, T., Feltzing, S., \& Lundstr{\"o}m, I.\ 2004, \textit{A\&A}, 415, 155

\bibitem[Boyajian et al.(2013)]{2013ApJ...771...40B} Boyajian, T.~S., von 
Braun, K., van Belle, G., et al.\ 2013, \textit{ApJ}, 771, 40 

\bibitem[Casagrande et 
al.(2010)]{2010A&A...512A..54C} Casagrande, L., Ram{\'{\i}}rez, I., Mel{\'e}ndez, J., Bessell, M., \& Asplund, M.\ 2010, \textit{A\&A}, 512, A54

\bibitem[Casagrande et al.(2012)]{2012ApJ...761...16C} Casagrande, L., 
Ram{\'{\i}}rez, I., Mel{\'e}ndez, J., \& Asplund, M.\ 2012, \textit{ApJ}, 761, 16

\bibitem[Chambers]{cha10} Chambers, J.~E.\ 2010, \textit{ApJ}, 724, 92 

\bibitem[Cochran et al.]{coc97} Cochran, W.~D., Hatzes, A.~P., Butler, R.~P., \& Marcy, G.~W.\ 1997, \textit{ApJ}, 483, 457 

\bibitem[Datson et al.(2012)]{2012MNRAS.426..484D} Datson, J., Flynn, C., 
\& Portinari, L.\ 2012, \textit{MNRAS}, 426, 484

\bibitem[Ecuvillon et 
al.(2006)]{2006A&A...445..633E} Ecuvillon, A., Israelian, G., Santos, N.~C., et al.\ 2006, \textit{A\&A}, 445, 633

\bibitem[Edvardsson et 
al.(1993)]{1993A&A...275..101E} Edvardsson, B., Andersen, J., Gustafsson, B., et al.\ 1993, \textit{A\&A}, 275, 101 

\bibitem[Fischer \& Valenti]{fv05} Fischer, D.~A., \& Valenti, J.\ 2005, \textit{ApJ}, 622, 1102 

\bibitem[Ghezzi et al.]{ghe10} Ghezzi, L., Cunha, K., Smith, V.~V., et al.\ 2010, \textit{ApJ}, 720, 1290 

\bibitem[Gonzalez]{gon97} Gonzalez, G.\ 1997, \textit{MNRAS}, 285, 403 

\bibitem[Gustafsson(1998)]{1998SSRv...85..419G} Gustafsson, B.\ 1998, \textit{Space Science Rev.}, 
85, 419 

\bibitem[Gustafsson(2008)]{2008PhST..130a4036G} Gustafsson, B.\ 2008, 
\textit{Physica Scripta Volume T}, 130, 014036 

\bibitem[Gustafsson et al.]{gus10} Gustafsson, B., Mel{\'e}ndez, J., Asplund, M., \& Yong, D.\ 2010, \textit{Ap\&SS}, 328, 185 

\bibitem[Kiselman et al.]{kis11} Kiselman, D., Pereira, T.~M.~D., Gustafsson, B., et al.\ 2011, \textit{A\&A}, 535, A14 

\bibitem[Laws \& Gonzalez (2001)]{law01} Laws, C., \& Gonzalez, G.\ 2001, \textit{ApJ}, 553, 405 

\bibitem[Lind et al.(2012)]{2012MNRAS.427...50L} Lind, K., Bergemann, M., 
\& Asplund, M.\ 2012, \textit{MNRAS}, 427, 50 

\bibitem[Mel{\'e}ndez et al.(2006)]{2006ApJ...641L.133M} Mel{\'e}ndez, J., 
Dodds-Eden, K., \& Robles, J.~A.\ 2006, \textit{ApJ} (Letters), 641, L133

\bibitem[Mel\'{e}ndez \& Ram\'{i}rez]{mel07} Mel\'{e}ndez, J. \& Ram\'{i}rez, I.  2007, \textit{ApJ} (Letters), 669, L89

\bibitem[Mel{\'e}ndez et al.]{mel09} Mel{\'e}ndez, J., Asplund, M., Gustafsson, B., \& Yong, D.\ 2009, \textit{ApJ} (Letters), 704, L66

\bibitem[Mel{\'e}ndez et 
al.(2010)]{2010A&A...522A..98M} Mel{\'e}ndez, J., Schuster, W.~J., Silva, J.~S., et al.\ 2010, \textit{A\&A}, 522, A98

\bibitem[Mel{\'e}ndez et al.]{mel12} Mel{\'e}ndez, J., Bergemann, M., Cohen, J.~G., et al.\ 2012, \textit{A\&A}, 543, A29 

\bibitem[Monroe, T. et al.]{} Monroe, T., Mel{\'e}ndez, J., Ram\'{i}rez, I. et al.\ 2013, \textit{ApJ} (Letters), submitted

\bibitem[Morbidelli et al. (2012)]{mor12} Morbidelli, A., Lunine, J.~I., O'Brien, D.~P., Raymond, S.~N., 
\& Walsh, K.~J.\ 2012, \textit{Ann. Rev. Earth Planet. Sci.}, 40, 251 

\bibitem[{\"O}nehag et al.]{one11} {\"O}nehag, A., Korn, A., Gustafsson, B., Stempels, E., \& Vandenberg, D.~A.\ 2011, \textit{A\&A}, 528, A85

\bibitem[Ram{\'{\i}}rez 
\& Mel{\'e}ndez(2005)]{2005ApJ...626..465R} Ram{\'{\i}}rez, I., \& Mel{\'e}ndez, J.\ 2005, \textit{ApJ}, 626, 465

\bibitem[Ram{\'{\i}}rez et 
al.(2007)]{2007A&A...465..271R} Ram{\'{\i}}rez, I., Allende Prieto, C., \& Lambert, D.~L.\ 2007, \textit{A\&A}, 465, 271 

\bibitem[Ram{\'{\i}}rez et al.]{ram09} Ram{\'{\i}}rez, I., Mel{\'e}ndez, J., \& Asplund, M.\ 2009, \textit{A\&A}, 508, L17 

\bibitem[Ram{\'{\i}}rez et al.(2010)]{ram10} Ram{\'{\i}}rez, I., Asplund, M., Baumann, P., Mel{\'e}ndez, J., \& Bensby, T.\ 2010, \textit{A\&A}, 521, A33

\bibitem[Ram{\'{\i}}rez et al.]{ram11} Ram{\'{\i}}rez, I., Mel{\'e}ndez, J., Cornejo, D., Roederer, I.~U., \& Fish, J.~R.\ 2011, \textit{ApJ}, 740, 76 

\bibitem[Ram{\'{\i}}rez et al.(2012)]{2012ApJ...752....5R} Ram{\'{\i}}rez, 
I., Michel, R., Sefako, R., et al.\ 2012, \textit{ApJ}, 752, 5

\bibitem[Schuler et al.]{sch11} Schuler, S.~C., Cunha, K., Smith, V.~V., et al.\ 2011, \textit{ApJ} (Letters), 737, L32 

\bibitem[Sousa et 
al.(2007)]{2007A&A...469..783S} Sousa, S.~G., Santos, N.~C., Israelian, G., Mayor, M., \& Monteiro, M.~J.~P.~F.~G.\ 2007, 
\textit{A\&A}, 469, 783 

\bibitem[Takeda 
\& Tajitsu(2009)]{2009PASJ...61..471T} Takeda, Y., \& Tajitsu, A.\ 2009, \textit{PASJ}, 61, 471

\bibitem[Udry \& Santos]{us07} Udry, S., \& Santos, N.~C.\ 2007, \textit{ARAA}, 45, 397 


\end{thebibliography}
\end{document}